\documentclass[pra,showpacs,twocolumn]{revtex4}%
\usepackage{amssymb}
\usepackage{amsfonts}
\usepackage{amssymb}
\usepackage{graphicx}
\usepackage{subfigure}
\usepackage{dcolumn}
\usepackage{bm}
\usepackage{amsmath}%
\providecommand{\U}[1]{\protect\rule{.1in}{.1in}}
\providecommand{\U}[1]{\protect\rule{.1in}{.1in}}
\begin{document}
\title{Effects of losses in the hybrid atom-light interferometer}
\author{Zhao-Dan Chen$^{1}$, Chun-Hua Yuan$^{1}$}
\email{chyuan@phy.ecnu.edu.cn}
\author{Hong-Mei Ma$^{1}$, Dong Li$^{1}$, L. Q. Chen$^{1}$}
\email{lqchen@phy.ecnu.edu.cn}
\author{Z. Y. Ou$^{1,2}$}
\author{Weiping Zhang$^{1}$}

\address{$^1$Quantum Institute for Light and Atoms, Department of Physics, East China
Normal University, Shanghai 200062, P. R. China}
\address{$^{2}$Department of Physics, Indiana University-Purdue University
Indianapolis, 402 North Blackford Street, Indianapolis, Indiana 46202, USA}

\begin{abstract}
Enhanced Raman scattering can be obtained by injecting a seeded light field
which is correlated with the initially prepared collective atomic excitation.
This Raman amplification process can be used to realize atom-light hybrid
interferometer. We numerically calculate the phase sensitivities and the
signal-to-noise ratios of this interferometer with the method of homodyne
detection and intensity detection, and give their differences between this two
methods. In the presence of loss of light field and atomic decoherence the
measure precision will be reduced which can be explained by the break of the
intermode decorrelation conditions of output modes.
\end{abstract}
\date{\today}

\maketitle

\section{Introduction}

Quantum parameter estimation is the use of quantum techniques to improve
measurement precision than purely classical approaches, which has been
received a lot of attention in recent years
\cite{Helstrom76,Holevo82,Caves81,Braunstein94,Braunstein96,Lee02,Giovannetti06,Zwierz10,Giovannetti04,Giovannetti11,Ou}%
. Interferometers can provide the most precise measurements. Recently,
physicists with the advanced Laser Interferometer Gravitational-Wave
Observatory (LIGO) observed the gravitational waves \cite{Abbott}. The
Mach--Zehnder interferometer (MZI) and its variants have been used as a
generic model to realize precise measurement of phase. In order to avoid the
vacuum fluctuations enter the unused port and are amplified in the
interferometer by the coherent light, Caves \cite{Caves81} suggested to
replace the vacuum fluctuations with the squeezed-vacuum light to reach a
sub-shot-noise sensitivity. Xiao et al. \cite{Xiao} and Grangier et al.
\cite{Grangier}\ have demonstrated the experimental results beyond the
standard quantum limit (SQL) $\delta\varphi=1/\sqrt{N}$ with $N$ number of
photons or other bosons. Due to overcoming the SQL and reaching the Heisenberg
limit (HL) $\delta\varphi=1/N$, it will lead to potential applications in high
resolution measurements. Therefore, many theoretical proposals and
experimental techniques are developed to improve the sensitivity \cite{Toth,
PezzBook, Demkowicz}. When the probe states made of correlated states, such as
the NOON states of the form $(|N\rangle_{a}|0\rangle_{b}+e^{i\phi_{N}%
}|0\rangle_{a}|N\rangle_{b})/\sqrt{2}$, the HL in the phase-shift measurements
can reach \cite{Dowling08,NOON}. But, high-$N$ NOON states is very hard to
synthesize. In the presence of realistic imperfections and noise, the ultimate
precision limit in noisy quantum-enhanced metrology was also studied
\cite{Dem09, Escher, Dem12,Berry13, Chaves, Dur, Kessler, Alipour}.

However, most of the current atomic and optical interferometers are made of
linear devices such as beam splitters and phase shifters. In 1986, Yurke et
al. \cite{Yurke86} introduced a new interferometer where two nonlinear beam
splitters take the place of two linear beam splitters (BSs) in the traditional
MZI. It is also called the SU(1,1) interferometer because it is described by
the SU(1,1) group, as opposed to SU(2) for BSs. The detailed quantum
statistics of the two-mode SU(1,1) interferometer was studied by Leonhardt
\cite{Leonhardt}. SU(1,1) phase states also have been studied theoretically in
quantum measurements for phase-shift estimation \cite{Vourdas,Sanders}. An
improved theoretical scheme of the SU(1,1) optical interferometer was
presented by Plick \emph{et al} \cite{Plick} who proposed to inject a strong
coherent beam to \textquotedblleft boost\textquotedblright\ the photon number.
Experimental realization of this SU(1,1) optical interferometer was reported
by different groups \cite{Jing11,Lett}. The noise performance of this
interferometer was analyzed \cite{Marino,Ou} and under the same phase-sensing
intensity condition the improvement of $4.1$ dB in signal-to-noise ratio was
observed \cite{Hudelist}. By contrast, SU(1,1) atomic interferometer also has
been experimentally realized with Bose-Einstein Condensates
\cite{Gross,Linnemann,Peise,Gabbrielli}. Gabbrielli \emph{et al.}
\cite{Gabbrielli} realized a nonlinear three-mode SU(1,1) atomic
interferometer, where the analogy of optical down conversion, the basic
ingredient of SU(1,1) interferometry, is created with ultracold atoms.

Collective atomic excitation due to its potential applications for quantum
information processing has attracted a great deal of interest
\cite{Duan,Polzik,Kuzmich}. Collective atomic excitation can be realized by
the Raman scattering. Initially prepared collective atomic excitation can be
used to enhance the second Raman scattering \cite{Chen09,Chen10,Yuan10}.
Subsequently, we proposed another scheme to enhance the Raman scattering using
the correlation-enhanced mechanism \cite{Yuan13}. That is, by injecting a
seeded light field which is correlated with the initially prepared collective
atomic excitation, the Raman scattering can be enhanced greatly, which was
also realized in experiment recently \cite{Chen15}. Such a photon-atom
interface can form an SU(1,1)-typed atom-light hybrid interferometer
\cite{ChenPRL15}, where the atomic Raman amplification processes replacing the
beam splitting elements in a traditional MZI \cite{Yurke86}. Different from
all-optical or all-atomic interferometers, the atom-light hybrid
interferometers depend on both atomic and optical phases so that we can probe
the atomic phases with optical interferometric techniques. The atomic phase
can be adjusted by magnetic field or Stark shifts. The atom-light hybrid
interferometer is composed of two Raman amplification processes. The first
nonlinear process generates the correlated optical and atomic waves in the two
arms and they are decorrelated by the second nonlinear process.

In this paper, we calculate the phase sensitivities and the SNRs using the
homodyne detection and the intensity detection. The differences between the
phase sensitivities and the SNRs are compared. The loss of light field and
atomic decoherence will degrade the measure precision. The effects of the
light field loss and atomic decoherence on measure precision can be explained
from the break of intermode decorrelation conditions.

Our article is organized as follows. In Sec. II, we give the model of the
hybrid atom-light interferometer, and in Sec. III we numerically calculate the
phase sensitivity and the SNR, and analyze and compare the conditions to
obtain the optimal phase sensitivity and the maximal SNR. In Sec. IV, the LCCs
of the amplitude quadrature and number operator are derived from the
light-atom coupling equations in the presence of light field loss and atomic
decoherence. The LCCs as a function of the transmission rate and the
collisional rate are calculated and analyzed. The loss of light field and
atomic decoherence will degrade the measure precision, which is explained from
the intermode decorrelation conditions break. Finally, in Sec. V we conclude
with a summary of our results.

\section{The Model of atom-light hybrid interferometer}

\begin{figure}[ptb]
\centerline{\includegraphics[scale=0.6,angle=0]{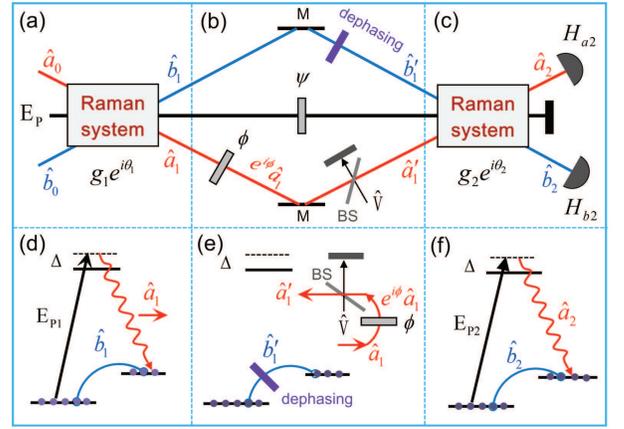}}\caption{(Color
online) (a) The intermode correlation between the Stokes field $\hat{a}_{1}$
and the atomic excitation $\hat{b}_{1}$ is generated by spontaneous Raman
process. $\hat{a}_{0}$ is the initial input light field. $\hat{b}_{0}$ is in
vacuum or an initial atomic collective excitation which can be prepared by
another Raman process or electromagnetically induced transparency process. (b)
During the delay time $\tau$, the Stokes field $\hat{a}_{1}$ will be subject
to the photon loss and evolute to $\hat{a}^{\prime}_{1}$ and the collective
excitation $\hat{b}_{1}$ will undergo the collisional dephasing to $\hat
{b}^{\prime}_{1}$. A fictitious beam splitter (BS) is introduced to mimic the
loss of photons into the environment. $\hat{V}$ is the vacuum. (c) After the
delay time $\tau$, the light field $\hat{a}^{\prime}_{1}$ and its correlated
atomic excitation $\hat{b}^{\prime}_{1}$ are used as initial seeding for
another enhanced Raman process. (d)-(f) The corresponding energy-level
diagrams of different processes are shown.}%
\label{fig1}%
\end{figure}

In this section, we review the different processes of the atom-light
interferometer \cite{ChenPRL15,Ma} as shown in Fig.~\ref{fig1}(a)-(c), where
two Raman systems replaced the BSs in the traditional MZI. Considering a
three-level Lambda-shaped atom system as shown in Fig. \ref{fig1}(d), the
Raman scattering process is described by the following pair of coupled
equations \cite{Raymer04}:%
\begin{equation}
\frac{\partial\hat{a}(t)}{\partial t}=\eta A_{P}\hat{b}^{\dag}(t),\text{
\ \ }\frac{\partial\hat{b}(t)}{\partial t}=\eta A_{P}\hat{a}^{\dag}(t),
\end{equation}
where $\eta$ is the coupling constant, and $A_{P}$ is the amplitude of the
pump field. The solution of above equation is%
\begin{equation}
\hat{a}(t)=u(t)\hat{a}(0)+v(t)\hat{b}^{\dag}(0),\text{ }\hat{b}(t)=u(t)\hat
{b}(0)+v(t)\hat{a}^{\dag}(0), \label{SRS1}%
\end{equation}
where $u(t)=\cosh(g)$, $v(t)=e^{i\theta}\sinh(g)$, $g=\left\vert \eta
A_{P}\right\vert t$, $e^{i\theta}=(A_{P}/A_{P}^{\ast})^{1/2}$, and $t$ is the
time duration of pump field $E_{P}$. We use different subscripts to
differentiate the two processes, where $1$ denotes the first Raman process
(RP1) and $2$ denotes the second Raman process (RP2). $t_{1}$ and $t_{2}$ are
the durations of the pump field $E_{P1}$ and $E_{P2}$, respectively.

After the first Raman process of the interferometer, the Stokes field $\hat
{a}_{1}$ and the atomic excitation $\hat{b}_{1}$ are generated as shown in
Fig.~\ref{fig1}(a). Then after a small delay time $\tau$, the second Raman
process of the interferometer takes place which is used as beams combination
as shown in Fig.~\ref{fig1}(c). During the small delay time $\tau$ shown in
Fig.~\ref{fig1}(b), the Stokes field $\hat{a}_{1}$ will be subject to the
photon loss and evolute to $\hat{a}_{1}^{\prime}$ . A fictitious BS is
introduced to mimic the loss of photons into the environment, then the light
field $\hat{a}_{1}^{\prime}$ is given by%
\begin{equation}
\hat{a}_{1}^{\prime}=\sqrt{T}\hat{a}_{1}(t_{1})e^{i\phi}+\sqrt{R}\hat{V},
\label{initial1}%
\end{equation}
where $T$ and $R$ are the transmission and reflectance coefficients with
$T+R=1$, and $\hat{V}$ is in vacuum. The collective excitation $\hat{b}_{1}$
will also undergo the collisional dephasing described by the factor
$e^{-\Gamma\tau}$, then $\hat{b}_{1}^{\prime}$ is%
\begin{equation}
\hat{b}_{1}^{\prime}=\hat{b}_{1}(t_{1})e^{-\Gamma\tau}+\hat{F},
\label{initial2}%
\end{equation}
where $\hat{F}=\int_{0}^{\tau}e^{-\Gamma(\tau-t^{\prime})}\hat{f}(t^{\prime
})dt^{\prime}$, and $\hat{f}(t)$ is the quantum statistical Langevin operator
describing the collision-induced fluctuation, and obeys $\langle\hat{f}%
(t)\hat{f}^{\dag}(t^{\prime})\rangle=2\Gamma\delta(t-t^{\prime})$ and
$\langle\hat{f}^{\dag}(t)\hat{f}(t^{\prime})\rangle=0$. Then $\langle\hat
{F}\hat{F}^{\dag}\rangle=1-e^{-2\Gamma\tau}$ guarantees the consistency of the
operator properties of $\hat{b}_{1}^{\prime}$.

Using Eqs. (\ref{SRS1})-(\ref{initial2}), the generated Stokes field $\hat
{a}_{2}$ and collective atomic excitation $\hat{b}_{2}$\ can be worked out:%
\begin{align}
\hat{a}_{2}(t_{2})  &  =\mathcal{U}_{1}\hat{a}_{1}(0)+\mathcal{V}_{1}\hat
{b}_{1}^{\dag}(0)+\sqrt{R}u_{2}\hat{V}+v_{2}\hat{F}^{\dag},\\
\hat{b}_{2}(t_{2})  &  =e^{-i\phi}[\mathcal{U}_{2}\hat{b}_{1}(0)+\mathcal{V}%
_{2}\hat{a}_{1}^{\dag}(0)]+\sqrt{R}v_{2}\hat{V}^{\dag}+u_{2}\hat{F},
\end{align}
where
\begin{align}
\mathcal{U}_{1}  &  =\sqrt{T}u_{1}u_{2}e^{i\phi}+e^{-\Gamma\tau}v_{1}^{\ast
}v_{2},\text{ }\mathcal{V}_{1}=\sqrt{T}v_{1}u_{2}e^{i\phi}\nonumber\\
&  +e^{-\Gamma\tau}u_{1}^{\ast}v_{2},\text{ }\mathcal{U}_{2}=e^{-\Gamma\tau
}u_{1}u_{2}e^{i\phi}\nonumber\\
&  +\sqrt{T}v_{1}^{\ast}v_{2},\text{ }\mathcal{V}_{2}=e^{-\Gamma\tau}%
v_{1}u_{2}e^{i\phi}+\sqrt{T}u_{1}^{\ast}v_{2}.
\end{align}

Next, we use the above results to calculate the phase sensitivity and the SNR,
and analyze and compare the conditions to obtain optimal phase sensitivity and
the maximal SNR.

\section{ Phase sensitivity and SNR}

\begin{figure}[ptbh]
\center
\subfigure[]{
\includegraphics[scale=0.4,angle=0]{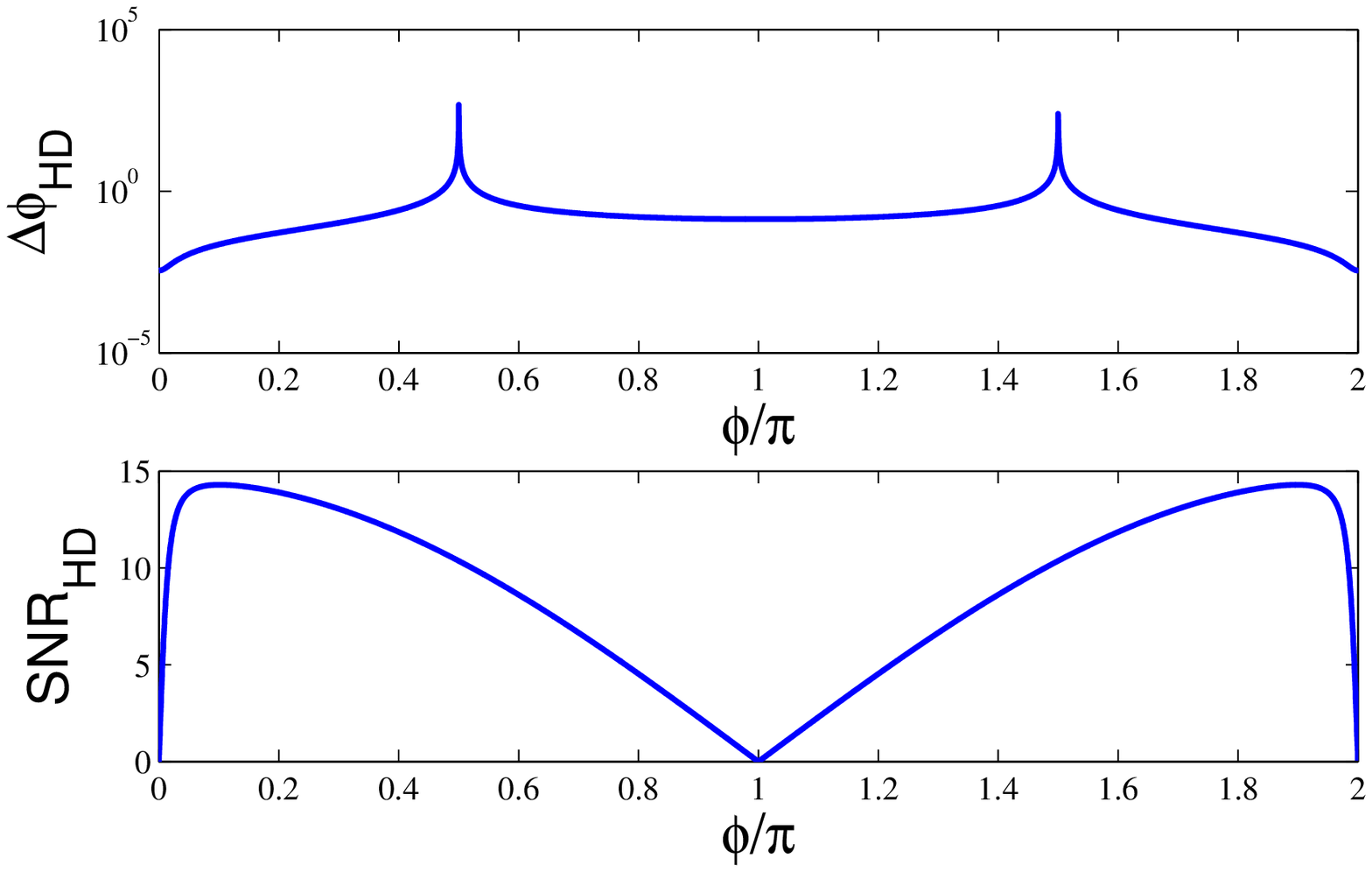}} \qquad\subfigure[]{
\includegraphics[scale=0.4,angle=0]{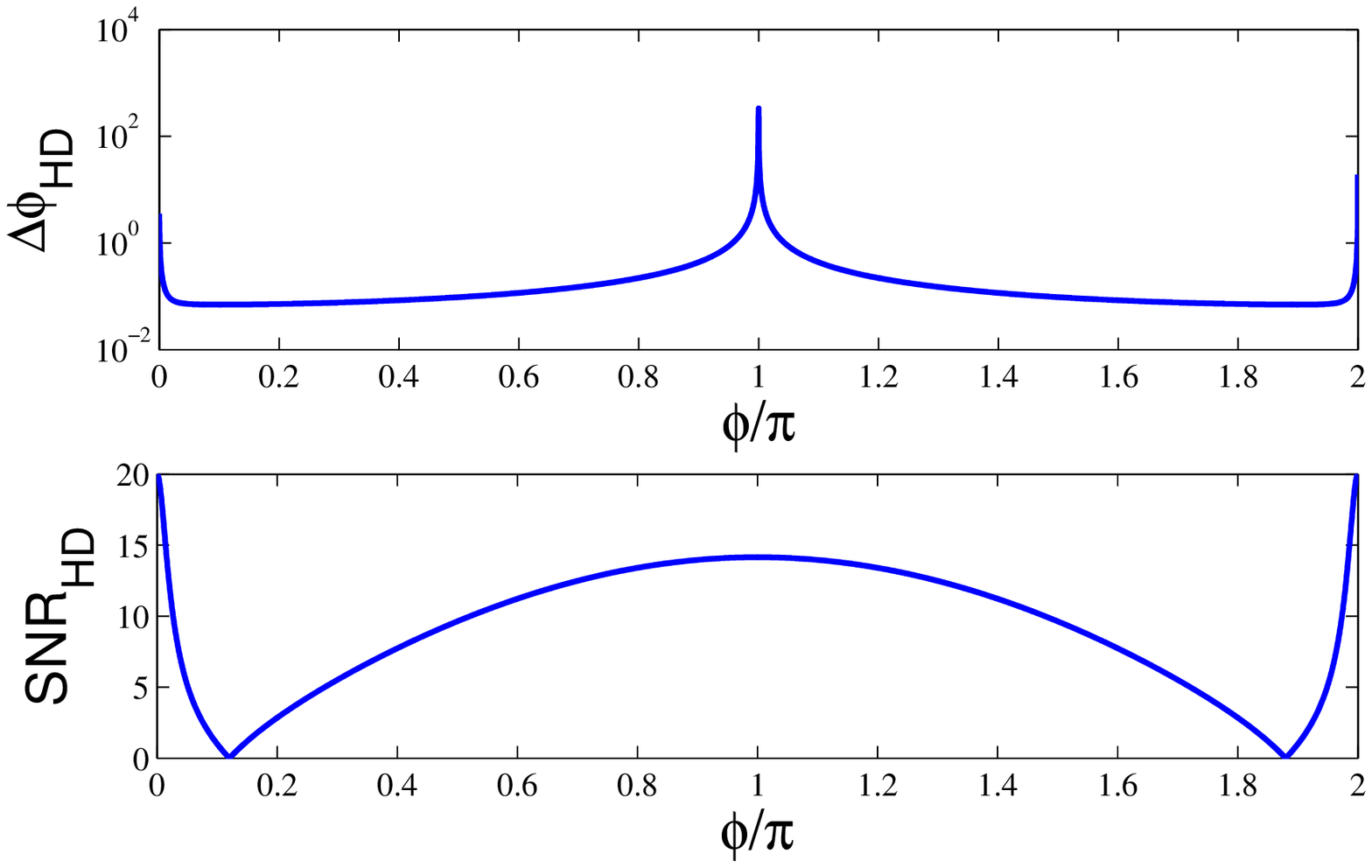}}\caption{(Color online) The
phase sensitivity $\Delta\phi_{\mathrm{HD}}$ and the $\mathrm{SNR_{HD}}$
versus the phase shift $\phi$ using the method of homodyne detection with (a)
$\theta_{\alpha}=\pi/2$; (b) $\theta_{\alpha}=0$. Parameters: $g=2$,
$|\alpha|=10$. }%
\label{fig2}%
\end{figure}

Phase can be estimated but cannot be measured because there is not a Hermitian
operator corresponding to a quantum phase \cite{Lynch}. In phase precision
measurement, the estimation of a phase shift can be done by choosing an
observable, and the the relationship between the observable and the phase is
known. The mean-square error in parameter $\phi$ is then given by the error
propagation formula \cite{Dowling08}:
\begin{equation}
\Delta\phi=\frac{\langle(\Delta\hat{O})^{2}\rangle^{1/2}}{\left\vert
\partial\langle\hat{O}\rangle/\partial\phi\right\vert }, \label{sens}%
\end{equation}
where $\hat{O}$ is the measurable operator and $\langle(\Delta\hat{O}%
)^{2}\rangle=\langle\hat{O}^{2}\rangle-\langle\hat{O}\rangle^{2}$. The
precision of the phase shift measurement is not the only parameter of concern.
We also need consider the SNR \cite{Kim,Ou,Barzanjeh}, which is given by%
\begin{equation}
\text{SNR}=\frac{\langle\hat{O}\rangle}{\langle(\Delta\hat{O})^{2}%
\rangle^{1/2}}.
\end{equation}

In current optical measurement of phase sensitivity, the homodyne detection
\cite{Ole,Li14,Barzanjeh} and the intensity detection \cite{Plick,Marino} are
often used. That is, the observables are the amplitude quadrature operator
$\hat{x}_{a2}=(\hat{a}_{2}+\hat{a}_{2}^{\dagger})/2$ and the number operator
$\hat{n}_{a2}=\hat{a}_{2}^{\dagger}\hat{a}_{2}$. For the balanced situation
that is $g_{1}=g_{2}=g$, and $\theta_{2}-\theta_{1}=\pi$. Firstly, we do not
consider the effect of loss on the generated Stokes field $\hat{a}_{2}$ and
atomic collective excitation $\hat{b}_{2}$. That is, $R=0$ and $\Gamma\tau=0$,
it reduced to the ideal lossless case and we have $\mathcal{U}_{1}%
=\mathcal{U}_{2}=\mathcal{U}=[\cosh^{2}ge^{i\phi}-\sinh^{2}g]$, $\mathcal{V}%
_{1}=\mathcal{V}_{2}=\mathcal{V}=\frac{1}{2}\sinh2g[e^{i\phi}-1]e^{i\theta
_{1}}$, where $\left\vert \mathcal{U}\right\vert ^{2}-\left\vert
\mathcal{V}\right\vert ^{2}=1$.

\begin{figure}[ptbh]
\center
\subfigure[]{
\includegraphics[scale=0.4,angle=0]{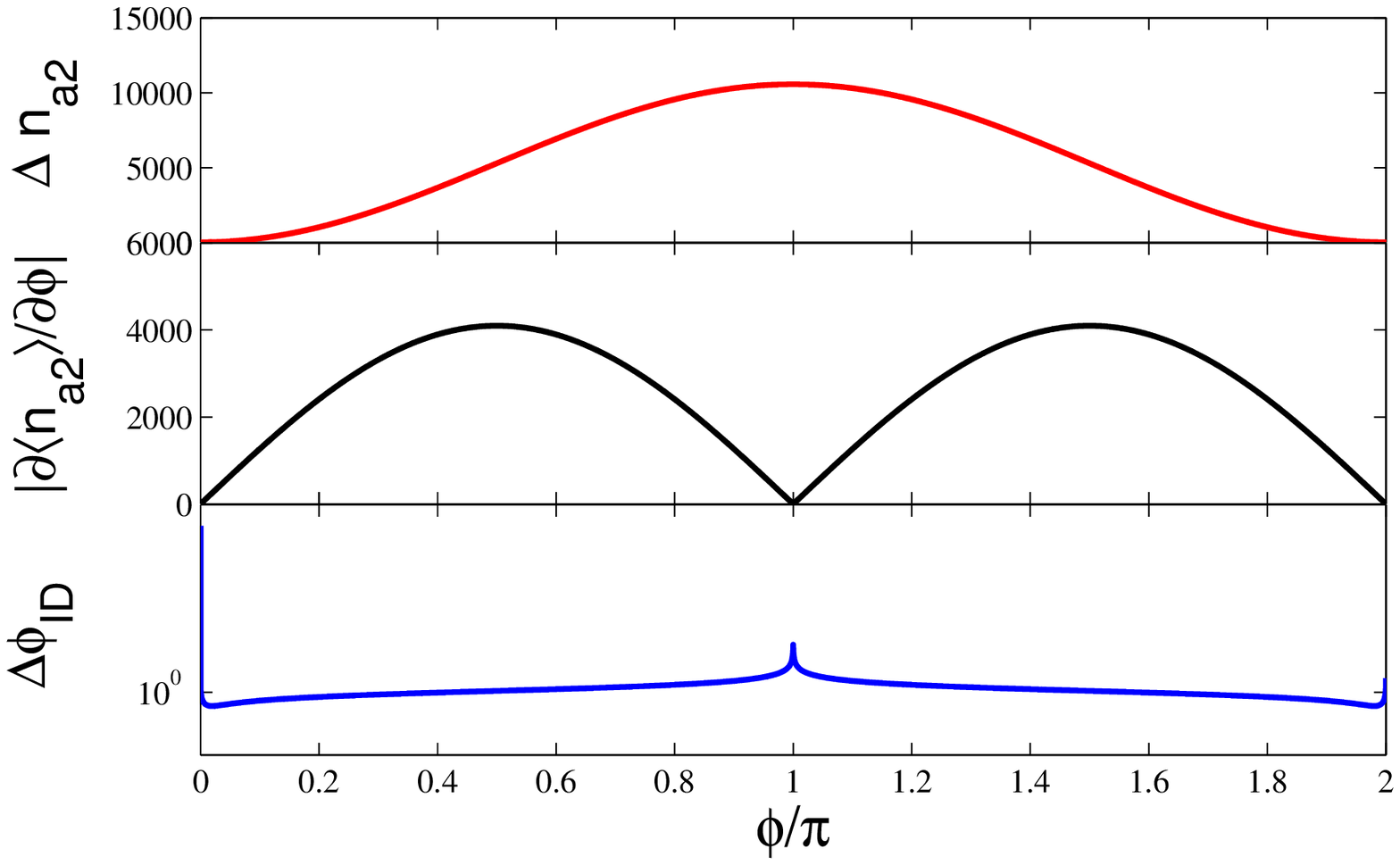}} \qquad\subfigure[]{
\includegraphics[scale=0.4,angle=0]{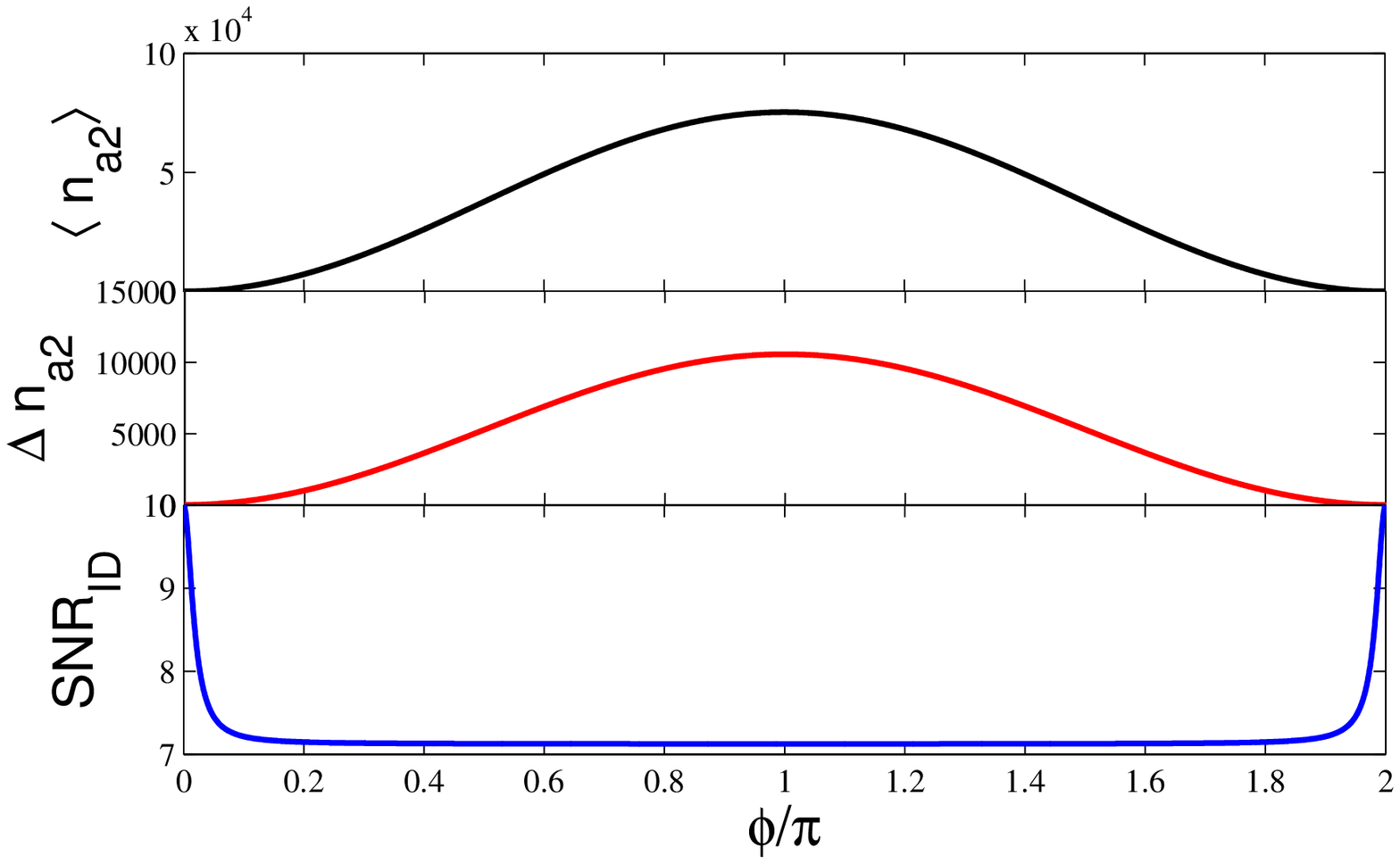}}\caption{(Color online) (a)
$\Delta{n}_{a2}$, $|\langle\partial\langle{n}_{a2}\rangle/\partial\phi
\rangle|$, and the phase sensitivity $\Delta\phi_{\mathrm{ID}}$; (b)
$\langle{n}_{a2}\rangle$, $\Delta{n}_{a2}$ and the $\mathrm{SNR_{ID}}$ versus
the phase shift $\phi$ using the method of intensity detection. Parameters:
$g=2$, $|\alpha|=10$. }%
\label{fig3}%
\end{figure}

\subsection{ Homodyne detection}

For a coherent light $|\alpha\rangle$ ($\alpha=\left\vert \alpha\right\vert
e^{i\theta_{\alpha}}$, $N_{\alpha}=\left\vert \alpha\right\vert ^{2}$) as the
phase-sensing field, using the amplitude quadrature operator $\hat{x}_{a2}$ as
the detected variable the phase sensitivity and the SNR are given by%
\begin{align}
\Delta\phi_{\text{HD}}  &  =\frac{\langle(\Delta\hat{x}_{a2})^{2}\rangle
^{1/2}}{\sqrt{N_{\alpha}}\cosh^{2}g\left\vert \sin(\phi+\theta_{\alpha
})\right\vert },\label{Sen}\\
\text{SNR}_{\text{HD}}  &  =\frac{\sqrt{N_{\alpha}}[\cosh^{2}g\cos(\phi
+\theta_{\alpha})-\sinh^{2}g\cos(\theta_{\alpha})]}{\langle(\Delta\hat{x}%
_{a2})^{2}\rangle^{1/2}}, \label{SNR}%
\end{align}
with%
\begin{equation}
\langle(\Delta\hat{x}_{a2})^{2}\rangle=\frac{1}{4}\left[  \cosh^{2}%
(2g)-\sinh^{2}(2g)\cos\phi\right]  ,
\end{equation}
where the subscript HD denotes the homodyne detection. The phase sensitivity
$\Delta\phi_{\text{HD}}$ and the SNR$_{\text{HD}}$ depend on $\phi$ and
$\theta_{\alpha}$, when $g$ and $\alpha$ take a certain values. From Eqs.
(\ref{Sen}) and (\ref{SNR}), both the $\Delta\phi_{\text{HD}}$ and the
SNR$_{\text{HD}}$ need that the term $\langle(\Delta\hat{x}_{a2})^{2}\rangle$
is minimal, which can be realized at $\phi=0$ and $\langle(\Delta\hat{x}%
_{a2})^{2}\rangle=1/4$ \cite{Scully}.

When $\phi=0$ and $\theta_{\alpha}=\pi/2$, we obtain the optimal phase
sensitivity and the worst SNR:
\begin{align}
\Delta\phi_{\text{HD}}  &  =\frac{1}{\sqrt{N_{\alpha}}}\frac{1}{2\cosh^{2}%
g},\\
\text{SNR}_{\text{HD}}  &  =0.
\end{align}
But when $\phi=0$ and $\theta_{\alpha}=0$ or $\pi$,\ the maximal
SNR$_{\text{HD}}$ is given by%
\begin{equation}
\text{SNR}_{\text{HD}}=2\sqrt{N_{\alpha}},
\end{equation}
and the sensitivity $\Delta\phi_{\text{HD}}$ is divergent. The phase
sensitivity $\Delta\phi_{\text{HD}}$ and the SNR$_{\text{HD}}$ of above two
different cases are shown in Figs. \ref{fig2}(a) and \ref{fig2}(b),
respectively. We find that at the optimal point $\phi=0$ and $\theta_{\alpha
}=\pi/2$ the sensitivity is high (i.e. $\Delta\phi$ small) and can beat the
SQL but the SNR$_{\text{HD}}$ is low. At the optimal point $\phi=0$ and
$\theta_{\alpha}=0$ the SNR$_{\text{HD}}$ is high, but the sensitivity is low.
Ideally, of course, we would like high sensitivity $\Delta\phi_{\text{HD}}$
and high SNR$_{\text{HD}}$ at the same optimal point.

\subsection{Intensity detection}

\begin{figure}[ptb]
\centerline{\includegraphics[scale=0.45,angle=0]{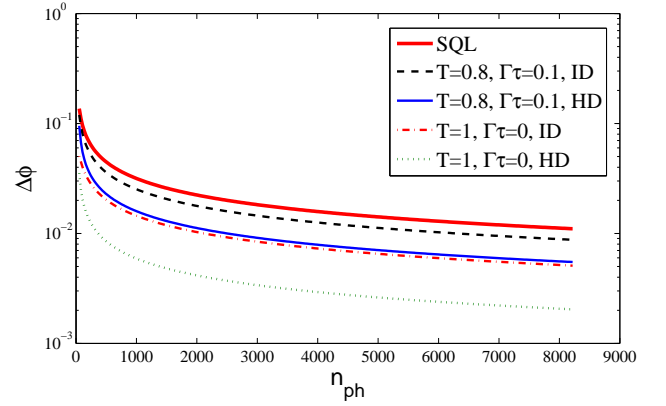}}\caption{(Color
online) The phase sensitivities $\Delta\phi$ versus the phase-sensing probe
number $n_{\mathrm{ph}}$. The optimal phase sensitivities $\Delta
\phi_{\text{HD}}$ and $\Delta\phi_{\text{ID}}$ are obtained at $\phi=0$ and
$\phi=0.062$, respectively. Parameter: $g=2$.}%
\label{fig4}%
\end{figure}

If we use $\hat{n}_{a2}$ ($=\hat{a}_{2}^{\dagger}\hat{a}_{2}$) as the
detection variable, for a coherent light $|\alpha\rangle$ ($\alpha=\left\vert
\alpha\right\vert e^{i\theta_{\alpha}}$, $N_{\alpha}=\left\vert \alpha
\right\vert ^{2}$) as the phase-sensing field, the phase sensitivity and the
SNR are given by%
\begin{align}
\Delta\phi_{\text{ID}}  &  =\langle(\Delta\hat{n}_{a2})^{2}\rangle^{1/2}%
\frac{2}{(N_{\alpha}+1)\sinh^{2}(2g)\left\vert \sin\phi\right\vert },\\
\text{SNR}_{\text{ID}}  &  =\frac{1}{\langle(\Delta\hat{n}_{a2})^{2}%
\rangle^{1/2}}[N_{\alpha}\left\vert \cosh^{2}g-\sinh^{2}ge^{-i\phi}\right\vert
^{2}\nonumber\\
&  +\frac{1}{2}\sinh^{2}(2g)(1-\cos\phi)],
\end{align}
where the subscript ID denotes the intensity detection, and%
\begin{align}
&  \langle(\Delta\hat{n}_{a2})^{2}\rangle=N_{\alpha}\left\vert \cosh
^{2}g-\sinh^{2}ge^{-i\phi}\right\vert ^{4}+\frac{1}{2}\left(  1+N_{\alpha
}\right) \nonumber\\
&  \times\sinh^{2}(2g)\left\vert \cosh^{2}g-\sinh^{2}ge^{-i\phi}\right\vert
^{2}(1-\cos\phi).
\end{align}

Different from the homodyne detection, the phase sensitivity $\Delta
\phi_{\text{ID}}$ and the SNR$_{\text{ID}}$ only depend on $\phi$ for given
$g$ and $N_{\alpha}$. Under the condition of $g=2$ and $N_{\alpha}=\sqrt{10}$,
the phase sensitivity $\Delta\phi_{\text{ID}}$ and the SNR$_{\text{ID}}$ as a
function of phase shift $\phi$\ are shown in Figs.~\ref{fig3}(a) and
\ref{fig3}(b), respectively. The best phase sensitivity $\Delta\phi
_{\text{ID}}$ and the maximal SNR$_{\text{ID}}$ ($=\sqrt{N_{\alpha}}$) are
obtained at $\phi\approx0.062$ and $\phi=0$, respectively. In Fig.
\ref{fig3}(a) at $\phi\approx0$ the slope $\left\vert \partial\langle\hat
{n}_{a2}\rangle/\partial\phi\right\vert $ is very small, as well in Fig.
\ref{fig3}(b) at $\phi=0$ the intensity of the signal $\langle\hat{n}%
_{a2}\rangle$ is low, but the noise is also low. It demonstrated that the
noise $\langle(\Delta\hat{n}_{a2})^{2}\rangle$ plays a dominant role. The best
phase sensitivity from the intensity detection\ is lower than it from the
homodyne detection, i.e., $\Delta\phi_{\text{ID}}>\Delta\phi_{\text{HD}}$. The
relation of maximal SNR from two detection methods is SNR$_{\text{HD}}%
=2$SNR$_{\text{ID}}$.

\subsection{Losses case}

If the presence of loss of light field and atomic decoherence, the precision
of the sensitivity and the SNR will be reduced \cite{Li14,Ma}. According to
the linear error propagation, the mean-square error in parameter $\phi$ is
given by%
\[
\Delta\phi=\frac{\langle(\Delta\hat{O})^{2}\rangle^{1/2}}{\left\vert
\partial\langle\hat{O}\rangle/\partial\phi\right\vert }.
\]
The slopes of the output amplitude quadrature operator $\hat{X}_{a2}$ and the
number operator $\hat{n}_{a2}=\hat{a}_{2}^{\dagger}\hat{a}_{2}$ are
respectively given by%
\begin{align}
\left\vert \frac{\partial\langle\hat{X}_{a2}\rangle}{\partial\phi}\right\vert
&  =\sqrt{TN_{\alpha}}\cosh^{2}g\left\vert \sin(\phi+\theta_{\alpha
})\right\vert ,\label{slope1}\\
\left\vert \frac{\partial\langle\hat{n}_{a}\rangle}{\partial\phi}\right\vert
&  =\frac{1}{2}\sqrt{T}e^{-\Gamma\tau}(N_{\alpha}+1)\sinh^{2}(2g)\left\vert
\sin(\phi)\right\vert .\label{slope2}%
\end{align}
The uncertainties of the output amplitude quadrature operator $\hat{X}_{a2}$
and the number operator$\ \hat{n}_{a2}$\ are given by%
\begin{align}
\langle(\Delta\hat{X}_{a2})^{2}\rangle &  =\frac{1}{4}[\sinh^{2}%
(2g)(T/2-\sqrt{T}e^{-\Gamma\tau}\cos\phi)\nonumber\\
&  +2e^{-2\Gamma\tau}\sinh^{4}g+\cosh(2g)],\\
\langle(\Delta\hat{n}_{a2})^{2}\rangle &  =\left\vert \mathcal{U}%
_{b}\right\vert ^{4}\left\vert \alpha\right\vert ^{2}+\left\vert
\mathcal{U}_{b}\mathcal{V}_{b}\right\vert ^{2}(1+\left\vert \alpha\right\vert
^{2})\nonumber\\
&  +R\cosh^{2}g(\left\vert \mathcal{U}_{b}\right\vert ^{2}\left\vert
\alpha\right\vert ^{2}+\left\vert \mathcal{V}_{b}\right\vert ^{2})\nonumber\\
+\sinh^{2}g &  [\left\vert \mathcal{U}_{b}\right\vert ^{2}(1+\left\vert
\alpha\right\vert ^{2})+R\cosh^{2}g](1-e^{-2\Gamma\tau}),
\end{align}
where%
\begin{align}
\left\vert \mathcal{U}_{b}\right\vert ^{2} &  =(\sqrt{T}\cosh^{2}%
g+e^{-\Gamma\tau}\sinh^{2}g)^{2}\nonumber\\
&  -2\sqrt{T}e^{-\Gamma\tau}\sinh^{2}g\cosh^{2}g(1+\cos\phi),\nonumber\\
\left\vert \mathcal{V}_{b}\right\vert ^{2} &  =\frac{1}{2}\sinh^{2}%
(2g)(T+e^{-2\Gamma\tau}-2\sqrt{T}e^{-\Gamma\tau}\cos\phi).
\end{align}
The subscript $b$ denotes the balanced condition when considering the losses case.

The phase sensitivities $\Delta\phi$ as a function of the phase-sensing probe
number $n_{\text{ph}}$\ is shown in Fig. \ref{fig4}. The thick solid line is
the SQL. The thin solid line and dotted line are sensitivities $\Delta
\phi_{\text{HD}}$ from homodyne detection with and without losses cases,
respectively. As well the dashed and dash-dotted lines are sensitivities
$\Delta\phi_{\text{ID}}$ from intensity detection with and without losses
cases, respectively. From Fig. \ref{fig4}, it is easy to see that the best
phase sensitivities $\Delta\phi_{\text{ID}}$ are larger than $\Delta
\phi_{\text{HD}}$ under the same condition. In the presence of the loss and
collisional dephasing ($T=0.8$, $\Gamma\tau=0.1$), the phase sensitivities
$\Delta\phi_{\text{HD}}$ and $\Delta\phi_{\text{ID}}$ can beat the SQL under
the balanced situation, which is very important for phase sensitivity measurement.

Next section, we explain the reason that the effects of the light field loss
and atomic decoherence on measure precision can be explained from the break of
intermode decorrelation conditions.

\section{The correlations of atom-light hybrid interferometer}

In this section, we use the above results to calculate the intermode
correlations of the different Raman amplification processes of the atom-light
interferometer as shown in Fig.~\ref{fig1}(a)-(c) \cite{ChenPRL15}. We also
study the effects of the loss of light field and the dephasing of atomic
excitation on the correlation. The intermode correlation of light and atomic
collective excitation can be described by the linear correlation coefficient
(LCC), which is defined as \cite{Gerry}%
\begin{equation}
J(\hat{A},\hat{B})=\frac{cov(\hat{A},\hat{B})}{\langle(\Delta\hat{A}%
)^{2}\rangle^{1/2}\langle(\Delta\hat{B})^{2}\rangle^{1/2}},
\end{equation}
where $cov(\hat{A},\hat{B})=(\langle\hat{A}\hat{B}\rangle+\langle\hat{B}%
\hat{A}\rangle)/2-\langle\hat{A}\rangle\langle\hat{B}\rangle$ is the
covariance of two-mode field and $\langle(\Delta\hat{A})^{2}\rangle
=\langle\hat{A}^{2}\rangle-\langle\hat{A}\rangle^{2}$, $\langle(\Delta\hat
{B})^{2}\rangle=\langle\hat{B}^{2}\rangle-\langle\hat{B}\rangle^{2}$.

The respective quadrature operators of the light and atomic excitation are
$\hat{x}_{a}=(\hat{a}+\hat{a}^{\dagger})/2$, $\hat{y}_{a}=(\hat{a}-\hat
{a}^{\dagger})/2i$, $\hat{x}_{b}=(\hat{b}+\hat{b}^{\dagger})/2$, and $\hat
{y}_{b}=(\hat{b}-\hat{b}^{\dagger})/2i$. After the first Raman scattering
process, the intermode correlations between the light field mode and the
atomic mode are generated. We start by injecting a coherent state
$|\alpha\rangle$ in mode $\hat{a}$, and a vacuum state in mode $\hat{b}$, the
LCC of quadratures are given by%
\begin{align}
J_{x1}(\hat{x}_{a1},\hat{x}_{b1})  &  =\cos\theta_{1}\tanh(2g_{1}%
),\label{eq4}\\
J_{y1}(\hat{y}_{a1},\hat{y}_{b1})  &  =-\cos\theta_{1}\tanh(2g_{1}),
\label{eq5}%
\end{align}
and the LCC of number operators $\hat{n}_{a1}$ $[=\hat{a}^{\dag}(t_{1})\hat
{a}(t_{1})]$ and $\hat{n}_{b1}$ $[=\hat{b}^{\dag}(t_{1})\hat{b}(t_{1})]$ is
given by%
\begin{equation}
J_{n1}(\hat{n}_{a1},\hat{n}_{b1})=\frac{(1+2\left\vert \alpha\right\vert
^{2})}{[4\coth^{2}(2g_{1})(\left\vert \alpha\right\vert ^{2}+\left\vert
\alpha\right\vert ^{4})+1]^{1/2}}. \label{eq6}%
\end{equation}
From Eqs.~(\ref{eq4})-(\ref{eq6}), the quadrature correlation LCCs
$J_{x1}(\hat{x}_{a1},\hat{x}_{b1})$\ and $J_{y1}(\hat{y}_{a1},\hat{y}_{b1})$
are independent on the input coherent state which is different from the number
correlation LCC $J_{n1}(\hat{n}_{a1},\hat{n}_{b1})$. Under $\theta_{1}\neq
\pi/2$, the LCCs $J_{x1}$ and $J_{y1}$ are opposite and not zero, which shows
the correlation exists. Due to their opposite intermode correlations, the
squeezing of quantum fluctuations is in a superposition of the two-modes,
i.e., $\hat{X}=(\hat{x}_{a}+\hat{x}_{b})/\sqrt{2}$, $\hat{Y}=(\hat{y}_{a}%
+\hat{y}_{b})/\sqrt{2}$ and $[\hat{X},\hat{Y}]=i/2$ \cite{Gerry}.

From Eq.~(\ref{eq6}) the number correlation LCC $J_{n1}$ is always positive so
long as$\ g\neq0$. If $\alpha=0$, that is vacuum state input, then
$J_{n1}(\hat{n}_{a1},\hat{n}_{b1})=1$, this maximal value shows the strong
intermode correlation and such states in optical fields are often called "twin
beams". For this vacuum state input case, the state of atomic mode and light
mode is similar to the two-mode squeezed vacuum state.

\begin{figure}[ptb]
\centerline{\includegraphics[scale=0.45,angle=0]{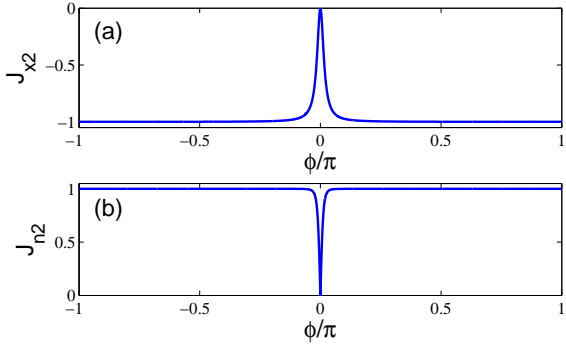}}\caption{(Color
online) The linear correlation coefficients (a) $J_{x2}$; (b) $J_{n2}$ as a
function of the phase shift $\phi$ for lossless case. Parameters: $\theta
_{1}=0$, $g=2$, $|\alpha|=10$. }%
\label{fig5}%
\end{figure}

\begin{figure}[ptb]
\centerline{\includegraphics[scale=0.4,angle=0]{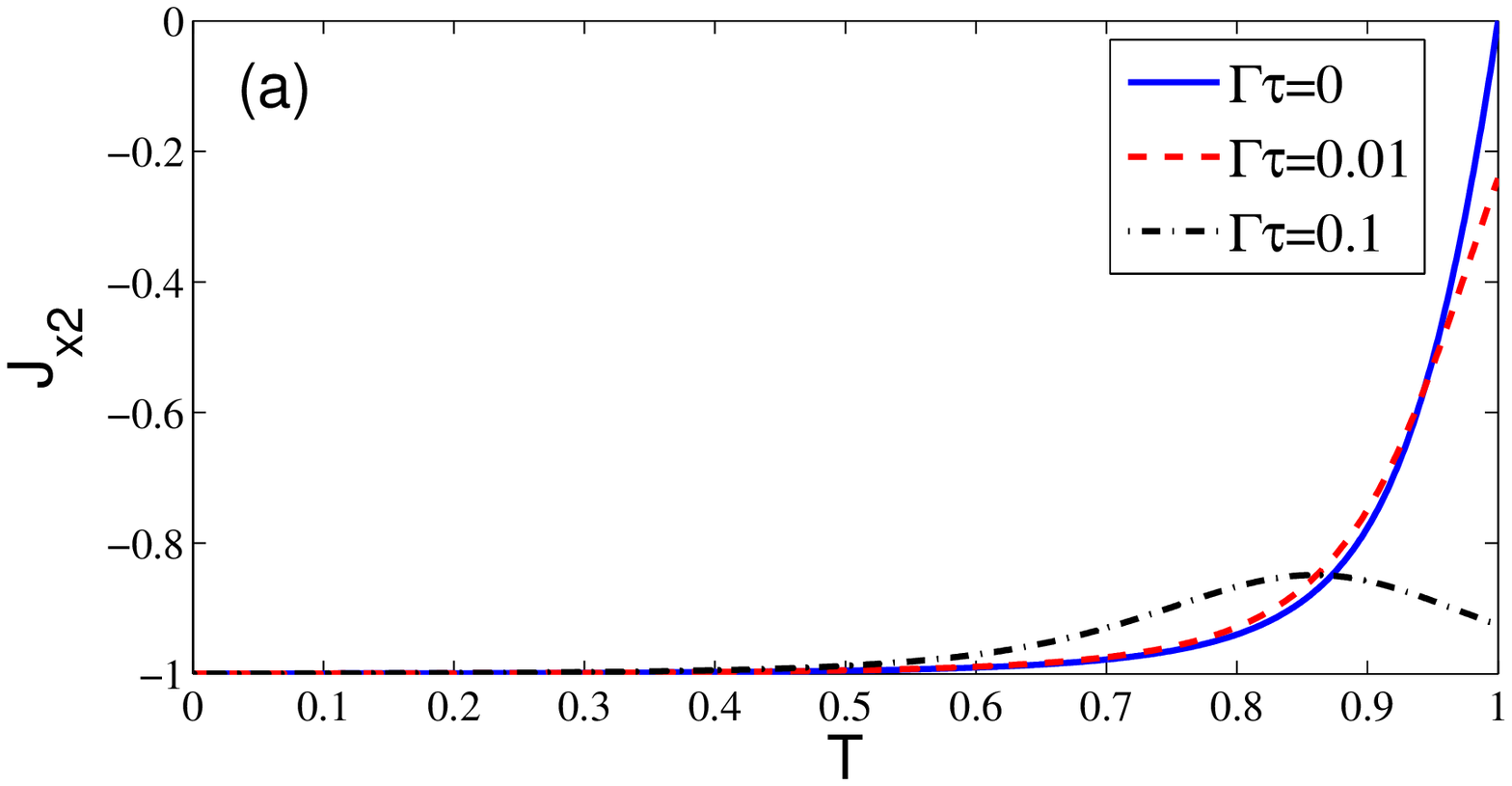}}
\centerline{\includegraphics[scale=0.4,angle=0]{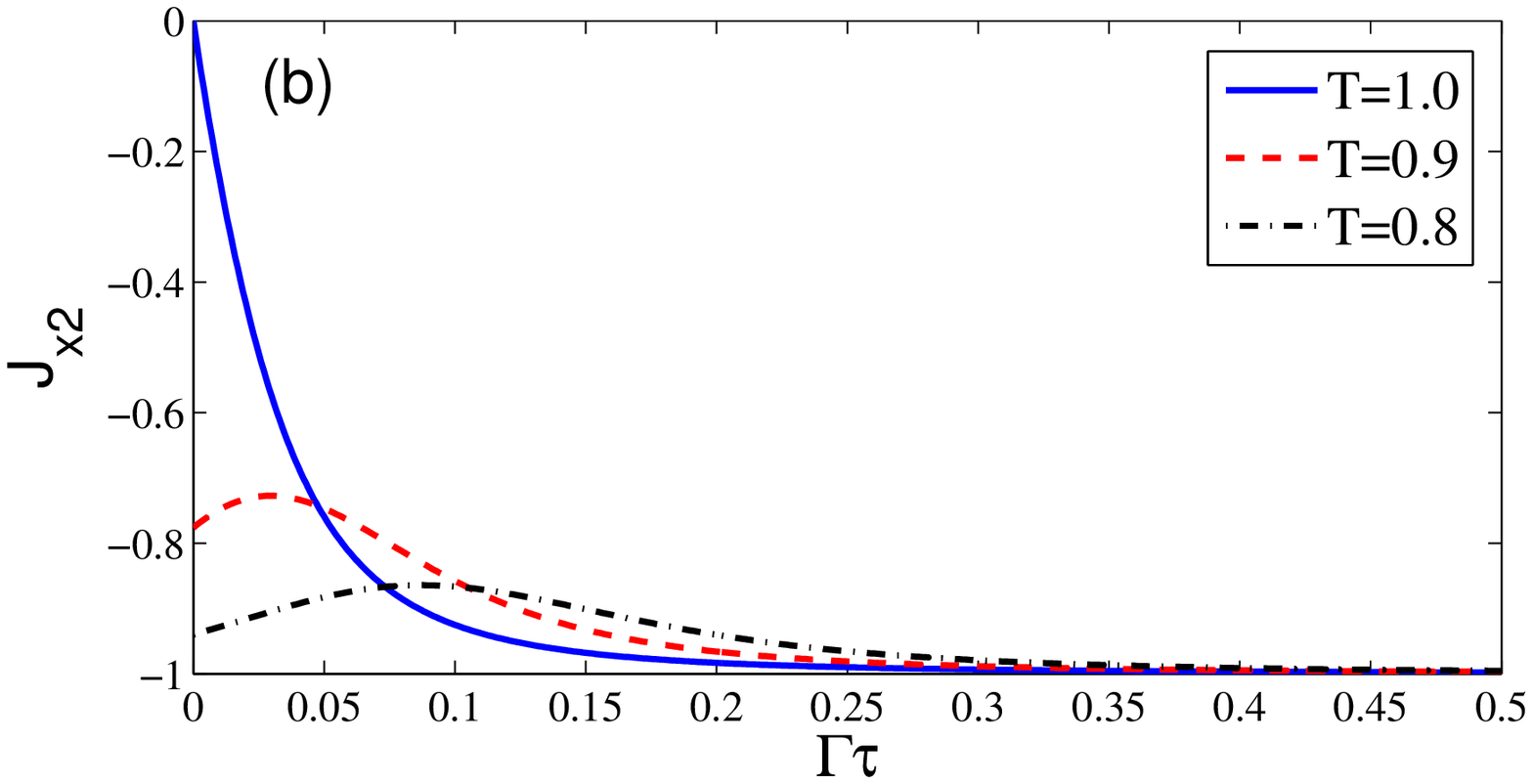}}\caption{(Color
online) The linear correlation coefficients $J_{x2}$ as a function of (a) the
transmission rate $T$; (b) the collisional rate $\Gamma\tau$. Parameters:
$g=2$, $|\alpha|=10$, $\theta_{\alpha}=\pi/2$ and $\phi=0$. }%
\label{fig6}%
\end{figure}

After the second Raman process of the interferometer, the LCC of quadratures
$J_{x2}(\hat{x}_{a2},\hat{x}_{b2})$ using the generated Stokes field $\hat
{a}_{2}$ and atomic collective excitation $\hat{b}_{2}$ can be worked out%
\begin{equation}
J_{x2}(\hat{x}_{a2},\hat{x}_{b2})=\frac{cov(\hat{x}_{a2},\hat{x}_{b2}%
)}{\langle(\Delta\hat{x}_{a2})^{2}\rangle^{1/2}\langle(\Delta\hat{x}_{b2}%
)^{2}\rangle^{1/2}},
\end{equation}
where
\begin{align}
&  cov(\hat{x}_{a2},\hat{x}_{b2})=\frac{1}{4}\mathrm{Re}[e^{-i\phi
}(\mathcal{V}_{1}\mathcal{U}_{2}+\mathcal{U}_{1}\mathcal{V}_{2})+u_{2}%
v_{2}\nonumber\\
&  \times(R+1-e^{-2\Gamma\tau})],\nonumber\\
&  \langle(\Delta\hat{x}_{a2})^{2}\rangle=\frac{1}{4}[\left\vert
\mathcal{U}_{1}\right\vert ^{2}+\left\vert \mathcal{V}_{1}\right\vert
^{2}+R\left\vert u_{2}\right\vert ^{2}+\left\vert v_{2}\right\vert
^{2}(1-e^{-2\Gamma\tau})],\nonumber\\
&  \langle(\Delta\hat{x}_{b2})^{2}\rangle=\frac{1}{4}[\left\vert
\mathcal{U}_{2}\right\vert ^{2}+\left\vert \mathcal{V}_{2}\right\vert
^{2}+R\left\vert v_{2}\right\vert ^{2}+\left\vert u_{2}\right\vert
^{2}(1-e^{-2\Gamma\tau})].
\end{align}
The LCC of number operators $J_{n2}(\hat{n}_{a2},\hat{n}_{b2})$ can also be
worked out%
\begin{equation}
J_{n2}(\hat{n}_{a2},\hat{n}_{b2})=\frac{cov(\hat{n}_{a2},\hat{n}_{b2}%
)}{\langle(\Delta\hat{n}_{a2})^{2}\rangle^{1/2}\langle(\Delta\hat{n}_{b2}%
)^{2}\rangle^{1/2}},
\end{equation}
where%
\begin{align}
&  cov(\hat{n}_{a2},\hat{n}_{b2})=\left\vert \mathcal{U}_{1}\mathcal{V}%
_{2}\right\vert ^{2}\left\vert \alpha\right\vert ^{2}+(1+\left\vert
\alpha\right\vert ^{2})\operatorname{Re}[\mathcal{U}_{1}^{\ast}\mathcal{U}%
_{2}\mathcal{V}_{1}\mathcal{V}_{2}^{\ast}]\nonumber\\
&  +(1-e^{-2\Gamma\tau})(R\left\vert u_{2}v_{2}\right\vert ^{2}+(1+\left\vert
\alpha\right\vert ^{2})\operatorname{Re}[e^{i\phi}\mathcal{U}_{1}^{\ast
}\mathcal{V}_{2}^{\ast}u_{2}v_{2}])\nonumber\\
&  +R\operatorname{Re}[e^{-i\phi}\mathcal{U}_{2}\mathcal{V}_{1}u_{2}^{\ast
}v_{2}^{\ast}]+R\left\vert \alpha\right\vert ^{2}\operatorname{Re}[e^{-i\phi
}\mathcal{U}_{1}\mathcal{V}_{2}u_{2}^{\ast}v_{2}^{\ast}],
\end{align}%
\begin{align}
&  \langle(\Delta\hat{n}_{a2})^{2}\rangle=\left\vert \mathcal{U}%
_{1}\right\vert ^{4}\left\vert \alpha\right\vert ^{2}+\left\vert
\mathcal{U}_{1}\mathcal{V}_{1}\right\vert ^{2}(1+\left\vert \alpha\right\vert
^{2})+R\left\vert \mathcal{V}_{1}u_{2}\right\vert ^{2}\nonumber\\
&  +R\left\vert \mathcal{U}_{1}u_{2}\right\vert ^{2}\left\vert \alpha
\right\vert ^{2}+\left\vert \mathcal{U}_{1}v_{2}\right\vert ^{2}%
(1-e^{-2\Gamma\tau})\left\vert \alpha\right\vert ^{2}\nonumber\\
&  +(\left\vert \mathcal{U}_{1}v_{2}\right\vert ^{2}+R\left\vert u_{2}%
v_{2}\right\vert ^{2})(1-e^{-2\Gamma\tau}),
\end{align}%
\begin{align}
&  \langle(\Delta\hat{n}_{b2})^{2}\rangle=\left\vert \mathcal{V}%
_{2}\right\vert ^{4}\left\vert \alpha\right\vert ^{2}+\left\vert
\mathcal{U}_{2}\mathcal{V}_{2}\right\vert ^{2}(1+\left\vert \alpha\right\vert
^{2})+R\left\vert \mathcal{U}_{2}v_{2}\right\vert ^{2}\nonumber\\
&  +R\left\vert \mathcal{V}_{2}v_{2}\right\vert ^{2}\left\vert \alpha
\right\vert ^{2}+\left\vert \mathcal{V}_{2}u_{2}\right\vert ^{2}%
(1-e^{-2\Gamma\tau})\left\vert \alpha\right\vert ^{2}\nonumber\\
&  +(\left\vert \mathcal{V}_{2}u_{2}\right\vert ^{2}+R\left\vert u_{2}%
v_{2}\right\vert ^{2})(1-e^{-2\Gamma\tau}).
\end{align}

Firstly, we do not consider the effect of loss on the generated Stokes field
$\hat{a}_{2}$ and atomic collective excitation $\hat{b}_{2}$. Under this ideal
and balanced conditions, the LCCs of quadratures\ and number operators\ are
respectively\ given by%
\begin{align}
J_{x2}  &  (\hat{x}_{a2},\hat{x}_{b2})=\frac{2\mathrm{Re}[\mathcal{VU}%
e^{i\phi}]}{\left\vert \mathcal{U}\right\vert ^{2}+\left\vert \mathcal{V}%
\right\vert ^{2}}\nonumber\\
&  =\frac{\sinh(2g)}{\cosh^{2}(2g)-\sinh^{2}(2g)\cos\phi}[\cosh^{2}%
g\cos(\theta_{1}+3\phi)\nonumber\\
&  +\sinh^{2}g\cos(\theta_{1}+\phi)-\cosh(2g)\cos(\theta_{1}+2\phi)],
\end{align}
and
\begin{align}
&  J_{n2}(\hat{n}_{a2},\hat{n}_{b2})=\frac{\left\vert \mathcal{UV}\right\vert
(1+2\left\vert \alpha\right\vert ^{2})}{(\mathcal{\bar{U}\bar{V}})^{1/2}%
}=(1+2\left\vert \alpha\right\vert ^{2})\nonumber\\
&  \times\left[  \frac{4\left[  1+\sinh^{2}(2g)(1-\cos\phi)\right]
^{2}(\left\vert \alpha\right\vert ^{2}+\left\vert \alpha\right\vert ^{4}%
)}{[1+\sinh^{2}(2g)(1-\cos\phi)]^{2}-1}+1\right]  ^{-1/2},
\end{align}
where $\mathcal{\bar{U}}=\left\vert \mathcal{U}\right\vert ^{2}\left\vert
\alpha\right\vert ^{2}+\left\vert \mathcal{V}\right\vert ^{2}(\left\vert
\alpha\right\vert ^{2}+1)$, $\mathcal{\bar{V}}=\left\vert \mathcal{V}%
\right\vert ^{2}\left\vert \alpha\right\vert ^{2}+\left\vert \mathcal{U}%
\right\vert ^{2}(\left\vert \alpha\right\vert ^{2}+1)$. When the phase shift
$\phi$ is 0, $\mathcal{V}$ is also equal to $0$, then the LCC $J_{x2}(\hat
{x}_{a2},\hat{x}_{b2})$ and $J_{n2}(\hat{n}_{a2},\hat{n}_{b2})$\ are $0$.
Under this condition, the RP2 will "undo" what the RP1 did. When the phase
shift $\phi$ is $\pi$, the LCC $J_{x2}(\hat{x}_{a2},\hat{x}_{b2})$ and
$J_{n2}(\hat{n}_{a2},\hat{n}_{b2})$\ are respectively given by%
\begin{align}
J_{x2}(\hat{x}_{a2},\hat{x}_{b2})  &  =-\tanh(4g)\cos(\theta_{1}),\\
J_{n2}(\hat{n}_{a2},\hat{n}_{b2})  &  =\frac{1+2\left\vert \alpha\right\vert
^{2}}{\sqrt{4\coth^{2}(2g_{1})(\left\vert \alpha\right\vert ^{2}+\left\vert
\alpha\right\vert ^{4})+1}}\nonumber\\
&  =J_{n1}(\hat{n}_{a1},\hat{n}_{b1}).
\end{align}
The LCCs $J_{x2}(\hat{x}_{a2},\hat{x}_{b2})$ and $J_{n2}(\hat{n}_{a2},\hat
{n}_{b2})$\ as a function of the phase shift $\phi$\ is shown in Fig.
\ref{fig5}. Due to the LCC $J_{x2}(\hat{x}_{a2},\hat{x}_{b2})$ is dependent on
$\theta_{1}$, the intermode correlation coefficients $J_{x2}(\hat{x}_{a2}%
,\hat{x}_{b2})$\ ranges between $-1$ and $0$ when $\theta_{1}=0$. The LCC
$J_{n2}(\hat{n}_{a2},\hat{n}_{b2})$ is positive, and $J_{n2}(\hat{n}_{a2}%
,\hat{n}_{b2})$\ ranges between $0$ and $1$.

\begin{figure}[ptb]
\centerline{\includegraphics[scale=0.4,angle=0]{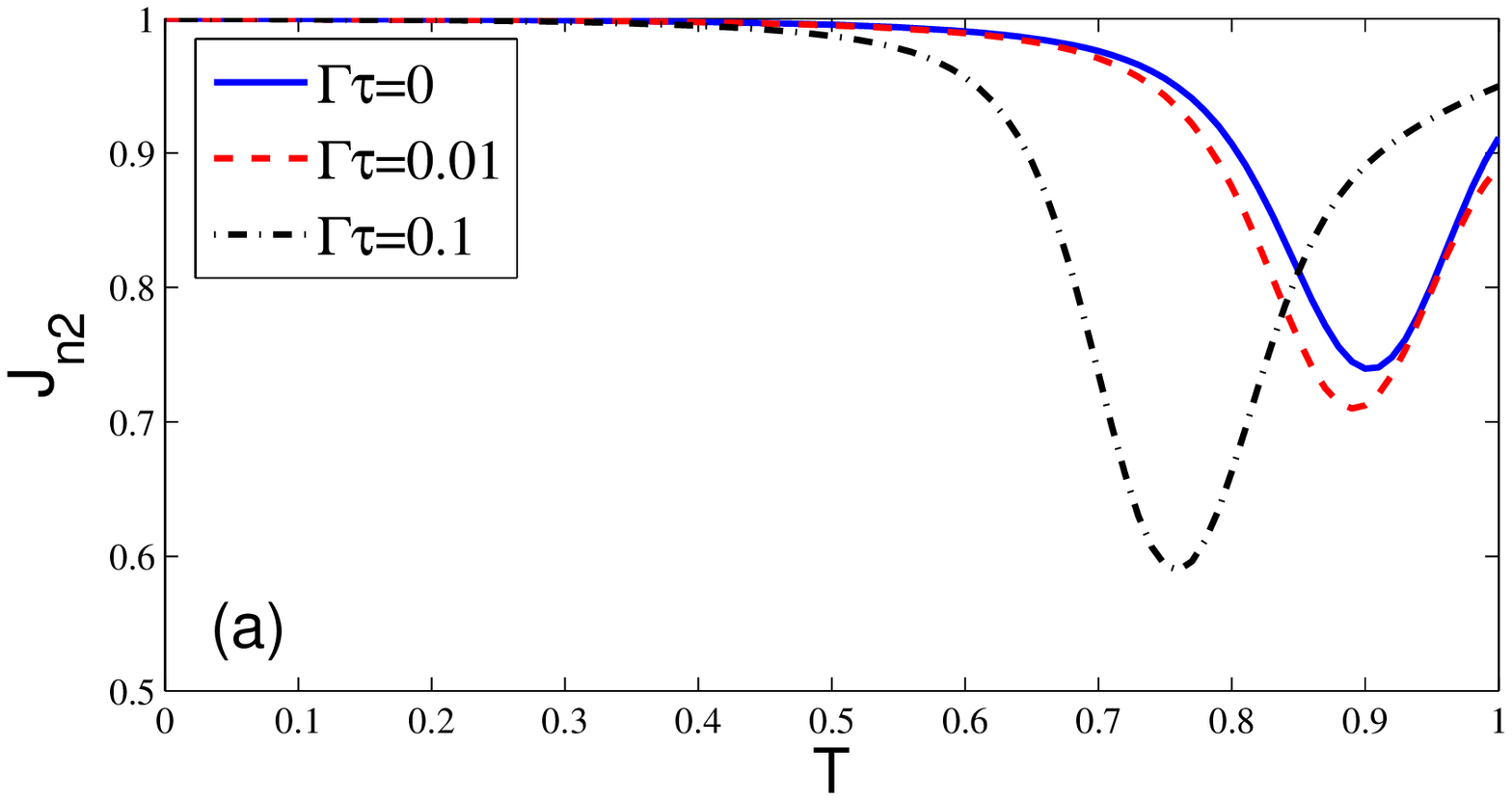}}
\centerline{\includegraphics[scale=0.4,angle=0]{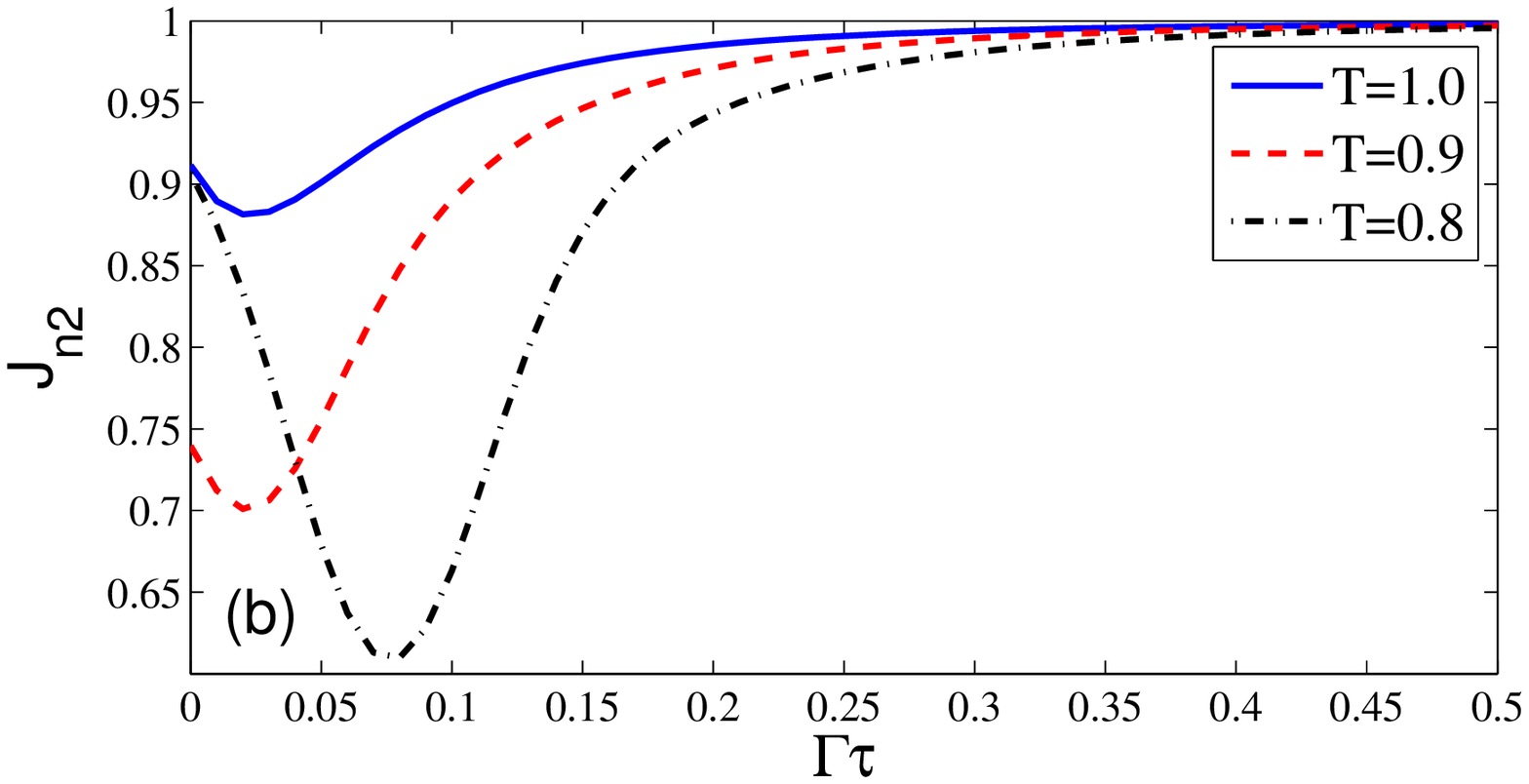}}\caption{(Color
online) The linear correlation coefficients $J_{n2}$ as a function of (a) the
transmission rate $T$ ; (b) the collisional dephasing rate $\Gamma\tau$, where
$g=2$, $|\alpha|=10$, and $\theta_{2}-\theta_{1}=\pi$, and $\phi=0.062$. }%
\label{fig7}%
\end{figure}

This decorrelation point ($\phi=0$) $J_{x2}(\hat{x}_{a2},\hat{x}_{b2})=0$ is
very important for atom-light hybrid interferometer using the homodyne
detection \cite{Ma}. At this point ($\phi=0$) the noise of output field
$[\langle(\Delta\hat{x}_{a2})^{2}\rangle=1/4]$ is the same as that of input
field and it is the lowest in our scheme as shown in Fig. \ref{fig2}. The
optimal phase sensitivity $\Delta\phi_{\text{HD}}$\ and the maximal
SNR$_{\text{HD}}$ are obtained at this point with different $\theta_{\alpha}$.
The LCC $J_{x2}$ as a function of the transmission rate $T$ and the
collisional dephasing rates $\Gamma\tau$ are shown in Fig. \ref{fig6}. With
the decrease of the transmission rate $T$ or the increase of $\Gamma\tau$, the
LCCs $J_{x2}$ is reduced and tend to $-1$. Due to large loss ($T$ small) or
large decoherence ($\Gamma\tau$ large) one arm inside the interferomter (the
optical field $\hat{a}_{1}^{\prime}$\ or the atomic excitation $\hat{b}%
_{1}^{\prime}$) is vanished, the decorrelation condition does not exist.
Therefore, the serious break of decorrelation condition will degrade the
sensitivity in the phase precision measurement.

This decorrelation point ($\phi=0$) $J_{n2}(\hat{n}_{a2},\hat{n}_{b2})=0$ is
also very important for the intensity detection. The low noise is dominant in
realizing the optimal sensitivity and the maximal SNR. At this point ($\phi
=0$), we can obtain the maximal SNR$_{\text{ID}}$. However, the slope
$\left\vert \partial\langle\hat{n}_{a2}\rangle/\partial\phi\right\vert $ is
equal to $0$ at this decorrelation point as shown in Fig. \ref{fig3}(a). In
Fig. \ref{fig3}(b) at nearby the decorrelation point, the noise is amplifed a
little and the optimal phase sensitivity $\Delta\phi_{\text{ID}}$ is obtained.
With the decrease of the transmission rate $T$ or the increase of $\Gamma\tau
$, the LCC $J_{n2}$ are reduced at first, then revive quickly, and finally
increase to $1$ as shown in Fig. \ref{fig7}. The behaviors of the two
detection methods are different, but both of their correlations eventually
tend to strong correlation due to the losses. Therefore, the serious break of
decorrelation condition will degrade the sensitivity in the phase precision measurement.

\section{Conclusions}

We gave out the phase sensitivities and the SNRs of the atom-light hybrid
interferometer with the method of homodyne detection and intensity detection.
Using the homodyne detection, for given input intensity $N_{\alpha}$ and
coupling intensity $g$ the optimal sensitivity $\Delta\phi_{\text{HD}}$\ and
the maximal SNR$_{\text{HD}}$ is not only dependent on the phase shift $\phi$
but also dependent on the phase $\theta_{\alpha}$ of the input coherent state.
We obtain that the sensitivity is low (i.e. $\Delta\phi_{\text{HD}}$ large)
when the SNR$_{\text{HD}}$ is high and vice versa because the optimal point
changes with $\theta_{\alpha}$. Using the intensity detection, the optimal
sensitivity $\Delta\phi_{\text{ID}}$ and the maximal SNR$_{\text{ID}}$ is only
dependent on the phase shift $\phi$ for given input intensity $N_{\alpha}$ and
coupling intensity $g$. Under the balanced condition, the maximal
SNR$_{\text{ID}}$ is obtained when the phase is $0$ and the optimal phase
sensitivity $\Delta\phi_{\text{ID}}$ is obtained when the phase is nearby $0$.
The loss of light field and atomic decoherence will degrade the sensitivity
and the SNR of phase measurement, which can be explained from the break of
decorrelation conditions.

\section{Acknowledgements}

This work was supported by the National Natural Science Foundation of China under Grant Nos.~11474095, ~11274118, ~11234003,~91536114 and~11129402, and is supported by the Innovation Program of the Shanghai
Municipal Education Commission (Grant No. 13ZZ036) and the Fundamental
Research Funds for the Central Universities.


\end{document}